\documentclass[useAMS,usenatbib]{mn2e}
\usepackage[final]{graphics}
\title[The emission distribution in RR Pictoris]
      {The emission distribution in RR Pictoris}
\author[L. Schmidtobreick, C. Tappert, I. Saviane]
       {L. Schmidtobreick$^{(1)}$\thanks{lschmidt@eso.org}, 
        C. Tappert$^{(2)}$\thanks{claus@gemini.cfm.udec.cl}, 
        and I. Saviane$^{(1)}$\thanks{isaviane@eso.org}\\
        $^{1}$European Southern Observatory, Casilla 19001, Santiago 19, Chile\\
        $^{2}$Grupo de Astronom\'{\i}a, Universidad de Concepci\'on,
              Casilla 160--C, Concepci\'on, Chile}
\begin{document}
\date{Accepted xxxx. Received xxxx; in original form xxxx}
\pagerange{\pageref{firstpage}--\pageref{lastpage}} \pubyear{2002}

\maketitle

\label{firstpage}

\begin{abstract}
We present time--resolved optical spectroscopy of the old nova RR\,Pic. 
Two emission lines (H$\alpha$ and He I)
are present in the observed part of the spectrum and both show strong
variability. H$\alpha$ has been used for Doppler tomography in order to
map the emission distribution in this system for the first time. 
The resulting map shows the emission from the disc as well as two 
additional emission 
sources on the leading and trailing side of the disc.
Furthermore we find evidence for the presence of either a disc--overflow
or an asymmetric outflow
from the binary with velocities up to $\pm 1200$\,km\,s$^{-1}$.
The origin of the outflow would
be the emission source on the leading side of the accretion disc.
\end{abstract}

\begin{keywords}
Physical data and processes: accretion, accretion discs -- 
stars: novae, cataclysmic variables -- individual: RR Pic.
\end{keywords}

\section{Introduction}
Cataclysmic Variables (CVs) are close, interacting binary systems, 
comprising a white dwarf receiving mass
from a Roche--lobe--filling late--type
star. In absence of strong magnetic fields, the mass transfer takes 
place via an accretion disc.
RR Pic belongs to the subclass of CVs known as classical novae, 
which have displayed
a thermonuclear runaway outburst.
The system has been discovered by \citet{spen31} at maximum light 
in 1925 and, although it was of slow type, it is supposed to be in its
quiescence state by now. The orbital period of 0.145025 days \citep{vogt75}
places it just above the period gap of CVs. 
Vogt found the lightcurve dominated by a very broad hump, often 
interrupted by superimposed minima. He explained this behaviour 
by an extended hot spot region with an inhomogeneous structure. 
\citet{haef+82} however, explained their own time--resolved photometric and 
polarimetric observations together with radial velocity variations of the
He\,II (4686\,\AA) emission line \citep{wyck+77} by 
suggesting the presence of an additional source of radiation opposite 
the hot spot. From high speed photometry, \citet{warn86} concluded that during 
the 1970s (about 50\,yr after outburst) structural changes have taken 
place in the system, resulting in a more isotropic  distribution of the 
emitted radiation. In addition, he has found evidence for a shallow, irregular
eclipse, showing RR\,Pic to be a high inclination system.
\citet{haef+91} confirmed the general change in the 
lightcurve of RR\,Pic as well as the presence of the eclipse, which they 
found to be very stable over several orbits. However, they point out that
the additional emission source on the trailing side is still necessary to
understand the phase relation between the radial velocity curve and the
photometric lightcurve. 

Apart from this, the question arose whether RR\,Pic is an intermediate polar.
Additional to the orbital period, \citet{kubi84} found a periodic modulation
in the optical with a 15\,min period. He interpreted this as the rotation of 
the white dwarf and concluded that RR\,Pic is an intermediate polar. 
\citet{haef+85} however, repeated the high time--resolved photometry on a 
longer time--scale and could not find any sign of this short period. 
Since no 15\,min period variation is found in X--ray measurements
\citep{beck+81} either,
they concluded that Kubiak's variation was more likely a transient event
in the disc rather than a rotating white dwarf. Also Warner's high-speed
photometry does not reveal any period other than the orbital one. Hence no 
evidence remains that RR\,Pic is an intermediate polar.

With a quiescence magnitude $m_V$ = 12\,mag, RR\,Pic is a comparatively
bright CV. Although it has been well observed photometrically, spectroscopic 
studies of the binary are sparse. In particular, no systematic investigation
of the line profile has been undertaken so far, which could provide information
on the accretion process in this system.
In this paper we present new time--resolved spectroscopic data on RR\,Pic 
and compare it
with photometric data from the literature. We establish a Doppler map of the
H$\alpha$ emission, study the isolated emission sources in the disc,
and discuss the models of RR\,Pic's accretion disc. Finally, we analyse
the high velocity wings of the H$\alpha$ line and discuss possible 
interpretations.

\section{Data}

\begin{figure}
\rotatebox{-90}{\resizebox{!}{8.5cm}{\includegraphics{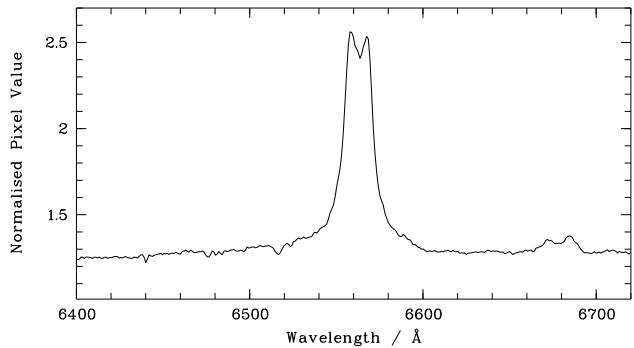}}}
\caption{\label{add} The 19 individual spectra have been added up  
and divided by the mean continuum to produce this average, normalised 
spectrum. Only the part around H$\alpha$ (6563\,\AA) and 
He\,I (6678\,\AA)
is plotted. }
\end{figure}
\begin{figure}
\resizebox{8.7cm}{!}{\includegraphics{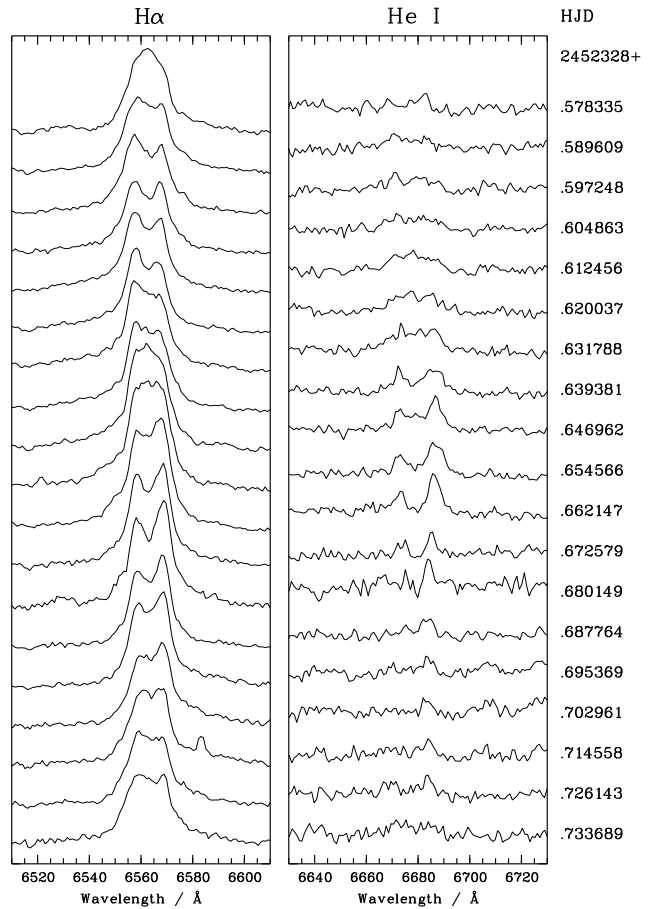}}
\caption{\label{spec_all} The time--resolved sequence of the two emission lines
H$\alpha$ (6563\,\AA) and He\,I (6678\,\AA)
is plotted. Note that the scaling of the two plots is different to ensure 
maximal visibility of both lines. The little feature on the red side of 
H$\alpha$ at epoch .714568 is an artefact due to a not properly subtracted
cosmic ray.}
\end{figure}

The time--resolved optical spectroscopy was obtained on 2002 February 23 with 
the Boller \& Chivens at the 1.52m ESO telescope on La Silla. 
Grating \#20 together with a 1.5 arcsec slit has been used 
to cover
the wavelength range from 5800 to 7500\,\AA\ with a spectral resolution 
of 2.45\,\AA\ FWHM. 
19 spectra were obtained, each
with an exposure time of 600\,s. In total, we cover 3.8\,h, which is slightly
more than a complete orbit. 
{\sc IRAF} has been used for the basic data reduction including 
BIAS subtraction, flat--fielding,
and wavelength calibration. No flux calibration has been performed. 
Two emission lines 
(H$\alpha$ and He\,I) are present in the observed part of the spectrum.
Both lines show a double peak profile in the average spectrum of all exposures
(Fig. \ref{add}) and strong
variability in time (Fig. \ref{spec_all}). 
H$\alpha$ has been used for further analysis in order to
map the emission distribution in the system, thus gaining insight
in the accretion structure and additional components. For this we used the
tomography code by \citet{spru98}, with a {\sc MIDAS} interface replacing
the original {\sc IDL} routines \citep{tapp+02}.

Note, that this technique makes use of the variability of the line
due to the binary rotation. We also expect a contribution  to the line
of the extended remnant shell of RR\,Pic which expands with a velocity 
around 400\,km\,s$^-1$ \citep{rosi+82}. However, since the shell is a 
circumbinary object, it is not effected by the binary rotation and any
contribution to the line profile is a stable one. It does therefore not effect
the here presented study.

\section{Discussion}
\subsection{Orbital period and zero phase}
\label{rvsection}
The radial velocities have been determined by applying the
double Gaussian method as described in \citet{shaf83}.
To evaluate the orbital period, two broad 
Gaussians have been fitted to each H$\alpha$ line. 
The Scargle algorithm \citep{scar82}
as implemented in MIDAS has been used to find the 
period in the radial velocities. As 
expected for data that covers one orbit only, 
the periodogram shows a broad maximum at $P=0.151$\,d. 
This maximum also includes the value of $P=0.1450255(2)$\,d, which
\citet{vogt75} has found from time--resolved photometry. 
We hence use this much more precise value
for the further analysis. 

To compare the observed spectral features with the
photometric lightcurves in the literature, it is essential to 
synchronise the respective phases. Unfortunately, Vogt's measurements are 
too far away in time 
from our observations for the accuracy of the period to yield an unambiguous 
zero phase. However, \citet{kubi84} later showed the constancy of the period
and even improved the value slightly, in addition \citet{warn86} and 
\citet{haef+91}
agree on the stability of the eclipse. 
Bringing these information together, we finally made use
of Warner's lightcurve S3464 from 
December 1984 \citep{warn86}, determined its HJD of the eclipse as 
zero point, and added Kubiak's period to create the ephemeris
\begin{equation}
{\rm eclipse} = {\rm HJD} 2450064.441(2) + 0.14502545(7) \cdot E,
\end{equation}
with E being the number of cycles.
These values are of sufficiently high accuracy to be extrapolated 
and compared to our observations.
Defining the zero phase $\phi_{\rm E} = 0$ as the time of the eclipse
we derive the orbital phase $\phi_{\rm E}$ for each observational point as
\begin{equation}
\label{phi}
\phi_E = \frac{\rm HJD_{obs} -  2450064.441}{0.14502545} + E
\end{equation}
Our observation starts $E=15612$ cycles after Warner's S3464 lightcurve.
For the zero phase we hence derive an uncertainty of $\sigma_0 = 0.003$\,d
or 0.02 phases. For comparison with Fig. \ref{spec_all}, the first 
observation (HJD = 2452328.578335) 
was obtained at phase $\phi_E = 7\cdot 10^{-5}$, the typical time between two 
exposures is equivalent to 0.053 phase units.

\subsection{Equivalent width}
\begin{figure}
\centerline{\rotatebox{-90}{\resizebox{!}{8.6cm}{\includegraphics{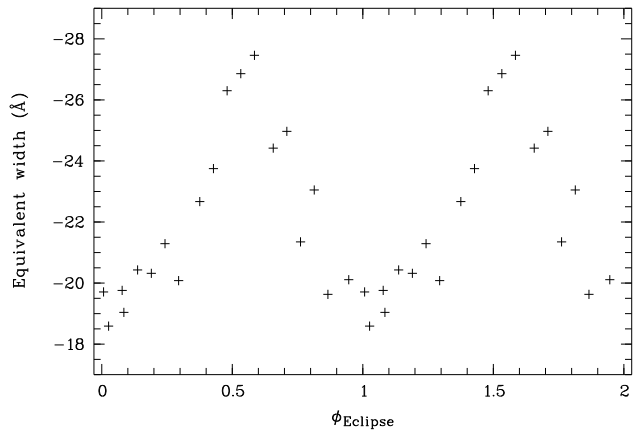}}}}
\caption{\label{eqw} Equivalent width of H$\alpha$ plotted against
photometric eclipse phase. Due to its breadth, the minimum at $\phi_E = 1$
is unlikely to represent an eclipse. }
\end{figure}
In Fig. \ref{eqw}, the equivalent width of the H$\alpha$--line has been 
plotted
against the orbital phase as derived from Equation \ref{phi}. 
A clear sinusoidal variation with a broad minimum around 1.0
and maximum around 0.5 orbits is detected. 
This behaviour agrees with measurements of the equivalent width 
by \citet{haef+91}.

A comparison with published lightcurves (\citet{haef+82}, \citet{warn86}, and 
\citet{haef+91}) shows that
the broad maximum in the lightcurves, which is partly 
eclipsed, does not coincide with maximal H$\alpha$ emission, but rather 
with the
broad minimum of the equivalent width. This indicates that it is 
probably caused by a region with enhanced continuum radiation,
which is hence more optically thick, and therefore suppresses the 
H$\alpha$ emission from this part of the disc. 
The maximal H$\alpha$ flux instead comes from a region situated on the
opposite side of the disc and coincides with the start of the minimum
at $\phi_{E}=0.7$ (named No\,1 by Warner) of the lightcurve.

Important for the later computation of the Doppler maps is the fact that, 
within the errors of the equivalent width, no eclipse feature is found 
in the line. 
Hence, all data can be used for the Doppler tomography.
Still, the variation of equivalent width requires a careful handling,
which is discussed in more detail in the corresponding section.
\subsection{Diagnostic Diagram}
\label{ddsection}
\begin{figure}
\centerline{\resizebox{8.7cm}{!}{\includegraphics{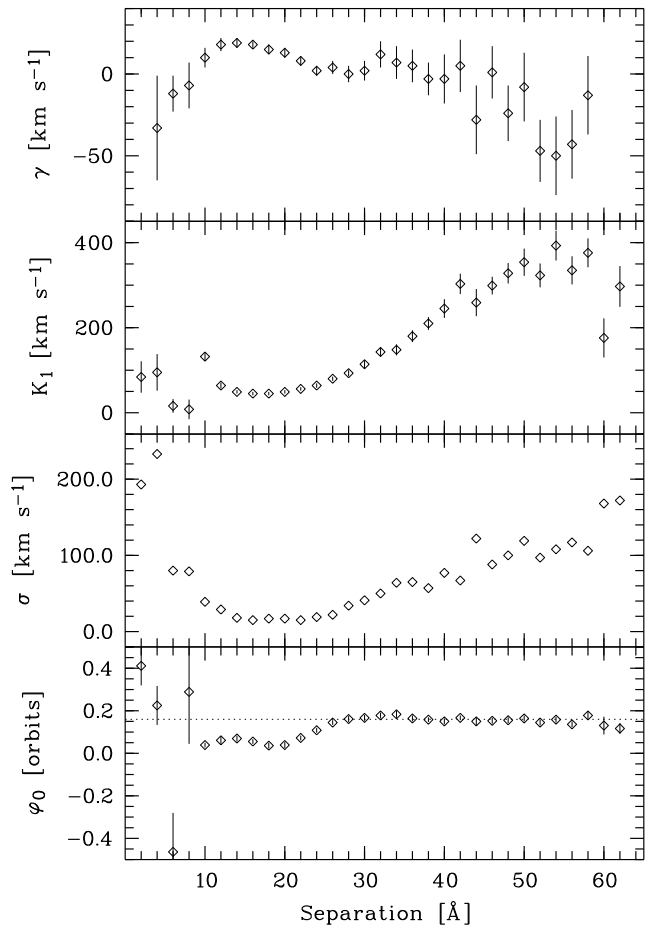}}}
\caption{\label{dd} Diagnostic diagram for RR\,Pic showing the parameters of
the radial velocity fit as function of the Gaussian separation.
The phases of the radial velocities were calculated
with respect to the eclipse ephemeris $\phi_E$. The $\phi_0$ plot
therefore gives the difference between $\phi_E$ and the zero point of the
radial velocity fit. The dotted line indicates the adopted correction
resulting from the diagnostic diagram.}
\end{figure}
While broad Gaussians have been sufficient to measure the orbital period,
greater care must be taken to determine those radial velocities, which
shall be used to study the line profile in more detail and to derive 
orbital parameters.
As the motion of the white dwarf should be best represented by the
extreme high--velocity line wings, which originate from the innermost part of
the disc, one aims for a maximum separation of the two Gaussians without
being contaminated by the continuum noise. This is usually achieved by a
diagnostic diagram \citep{shaf83} as shown in
Fig. \ref{dd}, 
where the parameters of the radial velocity fit,
\begin{equation}
v_r(\phi) = \gamma - K_1 \sin{2\pi \phi},
\end{equation}
are plotted as a function of the separation of the
two Gaussians. The FWHM of each Gaussian was chosen to be 2\AA. 
 The amplitude $K_1$ shows an extremely steep slope to
higher velocities with separation $d$ and reaches a maximum value of about 
350 km\,s$^{-1}$.
Although high values of $K_1$ might be expected from a high inclination 
system such as RR\,Pic, the value derived in this case is very improbable 
if not impossible. With $K_1 = 350$\,km\,s$^{-1}$ and
$P=0.145$\,d the distance for the white dwarf to the centre of mass 
is 0.9 solar radii. The maximum possible distance between
white dwarf and secondary for a CV with this period is roughly 1.6 solar radii.
Hence, the centre of mass would be situated nearer to the secondary than to
the primary, which would imply a mass ratio $M_2/M_1 > 1$, and does not allow 
for a stable mass transfer. Additionally,
we point out that the maximum values of $K_1$ are reached at very high  
separations, which are already far away from the line centre
in a region of very low S/N. We conclude that even at high separation,
the line is still affected by additional emission sources, which then must
be large and reach either deep into the central part of the disc or
do not belong to the disc at all, as discussed in section \ref{outflow}.

\begin{figure}
\centerline{\rotatebox{-90}{\resizebox{!}{8.6cm}{\includegraphics{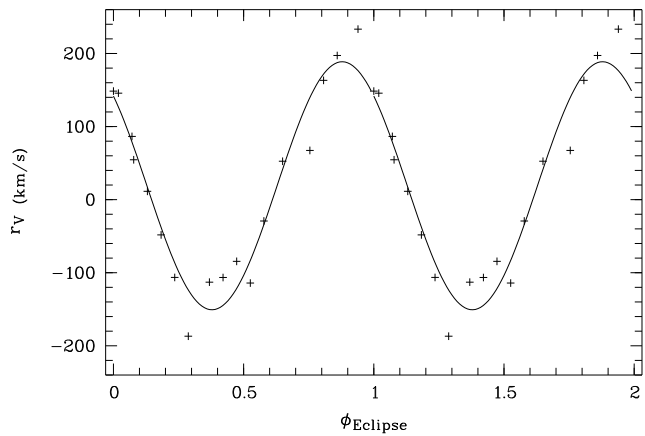}}}}
\caption{\label{rv34}
The radial velocities determined with the separation $d = 34$ of the
double Gaussian are plotted against the orbital phase. 
The line indicates the best sinusoidal fit to the data.
The semi--amplitude for this separation is $K_1 =  169.7\pm7$\,km\,s$^{-1}$.
The offset in radial velocities 
can be explained by the system velocity $\gamma$,
the offset between zero phase of radial velocity and photometric eclipse
is discussed in the text.}
\end{figure}

As a consequence of the steep increase of $K_1$, the ordinarily plotted
ratio $\sigma(K_1)/K_1$ stays nearly constant
and can not be used to indicate maximal possible separation.
We instead plot the error of the sinusoidal radial velocity fit $\sigma$
to give an idea of the uncertainties. The strong variation that starts
around $d\sim 34$ might indicate the point of optimal separation and
favours a semi--amplitude $K_1 \sim 170$\,km\,s$^{-1}$. 
In Fig. \ref{rv34}
the radial velocities for $d = 34$ are plotted against the orbital phase.

\begin{figure}
\centerline{\resizebox{8.3cm}{!}{\includegraphics{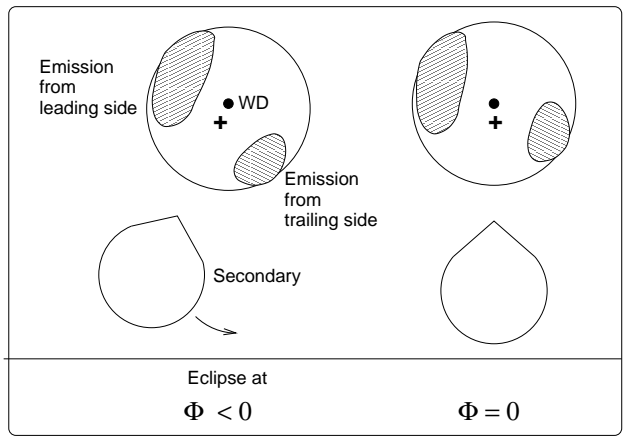}}}
\caption{\label{pic3}The sketch visualises the geometry of an eclipse
(left side)
before the superior conjunction of the white dwarf ($\phi_r = 0$, right side).
The cross marks the centre of rotation.
The eclipsed part has to be an emission source on the leading side of the disc.}
\end{figure}

After the initial distortion due to the double peaked line profile
\citep{tapp99} the zero phase settles at a value around 0.05
between separations $d \approx 10$ and 20.
A similar value has also been found by \citet{haef+82} who combined their
lightcurves with radial velocities measured by \citet{wyck+77}. 
This value for the zero phase is, however, very likely to be influenced by 
additional emission sources, which give a strong bias at low separations. 
Wyckhoff \& Wehinger state that they derived the radial velocities with
"classical methods", and therefore very probably used the total line for
their measurements. Hence, their zero phase can be assumed to be biased
in the same way as our values at small d.

For higher separations ($d\sim30$ and above), which are 
usually believed to more likely 
reflect the motion of the white dwarf, the zero phase
reaches $\phi_0 = 0.16$ and stays almost constant afterwards.
It therefore very likely corresponds to the 
superior conjunction of the white dwarf and
gives an unambiguous input value  
for the subsequent Doppler mapping. 
Note that, since the orbital phase was selected in such a way that 
$\phi_E =0$ for the eclipse (see Equation \ref{phi}), 
the zero phase $\phi_0$ indicates the difference between the phase of the 
eclipse and the zero phase of radial velocities; 
i.e. the eclipse appears 0.16 orbits before
the superior conjunction of the white dwarf (see also Fig. \ref{rv34}).
This can only be explained with the assumption 
that not the disc as a whole is eclipsed, but rather an emission 
region on the leading side of the disc (Fig. \ref{pic3}).  

\subsection{Doppler map}
\begin{figure}
\centerline{\resizebox{8.9cm}{!}{\includegraphics{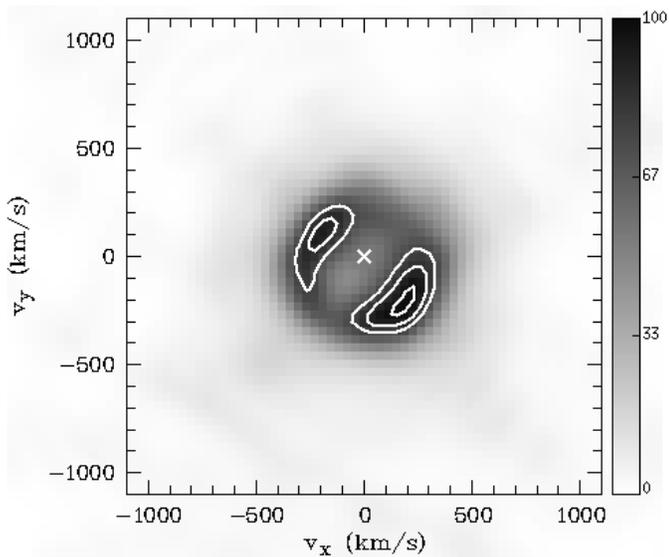}}}
\caption{\label{dop} Doppler map of RR\,Pic showing the distribution of
H$\alpha$--sources in velocity coordinates. The cross marks the centre of
rotation, which corresponds to the centre of gravity of the system. The map
is oriented in such a way that the phase angle referring to the zero point of 
radial velocities is zero towards the top and increases clockwise.
The contour lines refer to values 75, 85, and 95 and indicate the 
relative brightness of the emission sources. }
\end{figure}

For the interpretation of the phase dependent line profiles,
the Doppler tomography method as developed by \citet{mars+88}
has been applied.
A Doppler map $I(v_x,v_y)$ displays the flux emitted by gas moving
with the velocity $(v_x,v_y)$.
This deprojection is achieved due to the rotation of the binary system,
as the projection of the velocity $(v_x,v_y)$ follows a sinusoidal
radial velocity curve. Hence, the line profiles can be transformed into
a map $I(v_x,v_y)$ without any assumptions about the actual velocity field.

To correct for the variation of the equivalent width,
the input spectra have been normalised by the H$\alpha$ emission line flux.
The intensity values in the Doppler map are hence to be interpreted as 
relative flux values only. 
Furthermore, we corrected
the orbital phase of the input spectra by the zero value $\phi_0 = 0.16$.
\begin{figure}
\rotatebox{-90}{\resizebox{!}{8.5cm}{\includegraphics{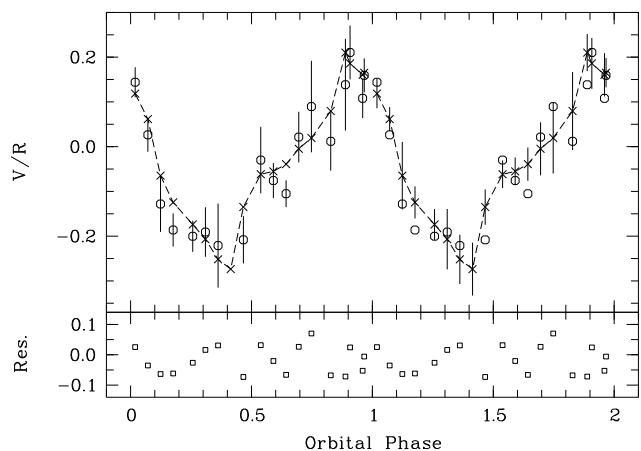}}}
\caption{\label{vrn}
$V/R$ and its residuals is  plotted against
the orbital phase referring to radial velocity zero point.
Error bars have been determined by a Monte Carlo simulation.
Two orbits are plotted, the original data ($\circ$)
with error bars between phases 0.0 and 1.0, and the reconstructed ($\diamond$)
one with error bars between phases 1.0 and 2.0. The latter points have 
additionally been connected via dashed lines. }
\end{figure}

The final Doppler map is given in Fig. \ref{dop} and 
shows the distribution of H$\alpha$ emission sources
in velocity coordinates. Apart from the disc itself, which is clearly visible,
two significant additional emission sources can be seen.
The orientation of the map is chosen in such a way that the phase angle 
$\phi_r$ with respect to the 
zero point of radial velocities, is zero towards the top and increases 
clockwise.
The smaller of the two emission sources is situated at a phase
angle $\phi_r \approx 0.8$, a  place that
is usually identified with emission from the hot spot.
The stronger emission, however, arises on the opposite side, $180^\circ$ away,
at $\phi_r \approx 0.3$. 

The quality of the Doppler map can be judged in several ways.
In this paper a quantitative approach is made to compare the
shape of the emission line in the original and reconstructed spectrum.
The asymmetry of the line profile can be measured by a V/R plot.
Here, the line profile is divided in two halves,
and V/R defined as logarithm of the ratio of the fluxes in the 
blue (\underline{v}iolet) and \underline{r}ed half \citep{tapp+02}.
In Fig. \ref{vrn}, $V/R$ of the original and reconstructed
line profile is plotted against the orbital phase. The residuals
given below show no sign of periodicity, hence indicating the good quality 
of the fit. 

\subsection{High velocity wings}
\label{outflow}

\begin{figure}
\resizebox{8.6cm}{!}{\includegraphics{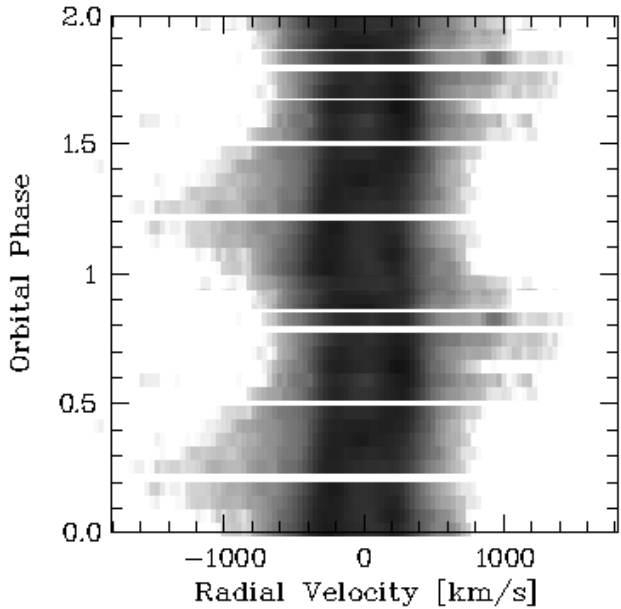}}
\caption{\label{swave} The observed H$\alpha$ emission line profile
of RR\,Pic has been plotted against the phase. The spectra have been
flux normalised as described in the text and the wavelength transformed
into radial velocities with respect to $\lambda_0 = 6562.8$\,\AA.
Intensities are logarithmically colour coded to favour the display
of the faint outer wings. The little enhancement at phase 0.83 and 
900\,km\,s$^{-1}$ is due to a not properly subtracted cosmic ray.}
\end{figure}
In Fig. \ref{swave} the changing line profiles of the flux normalised
H$\alpha$ emission line are displayed as an image with
emphasis on the faint outer parts of the line.
One can clearly see that the line wings reach up to velocities of 
1200\,km\,s$^{-1}$ and vary with the orbital phase.

The classical explanation for this feature is that it is caused 
by gas circling the white dwarf at velocities of about 1200\,km\,s$^{-1}$.
However, there are some problems with this picture. According to 
Fig. \ref{swave}, the gas not
only has these high velocities, but also changes its velocity by the same
magnitude with phase; it is completely on the side of negative 
velocities at phase $\phi_r = 0.25$ and on the side of positive 
velocities at $\phi_r = 0.75$. As the high velocity gas is 
supposed to be in the inner part
of the disc, this indicates a semi--amplitude $K_1 \approx 1000$\,km\,s$^{-1}$, which 
is a very unlikely value, especially as the distance of the white dwarf from 
the centre of mass in the Doppler map indicates a much smaller $K_1$.

Similar high velocity wings have been found in the Balmer lines of the
SW\,Sex--type systems, e.g. LS\,Peg \citep{tayl+99}. 
They explain the wings 
and their variation with a modified disc--overflow model similar to
the one developed by \citet{hell+94}. There are
certainly similarities between RR\,Pic and SW\,Sex--type stars, as
the orbital period is between 3\,h and 4\,h and the radial velocity phase
lags behind the eclipse phase (which we have explained by an eclipse of
the emission source on the leading side). However, the dominant absorption
feature around phase 0.5, which is typical for SW\,Sex--type stars and
probably related to the disc--overflow is not present in RR\, Pic.
Also, if all accreted material were deflected over the disc, no
hot spot could be present. However, some SW\,Sex--type stars are known
to show emission sources on the trailing side of their discs. Hence the
possibility of part of the accretion stream flowing over the disc can not
be ruled out for RR\,Pic.

\begin{figure}
\resizebox{8.6cm}{!}{\includegraphics{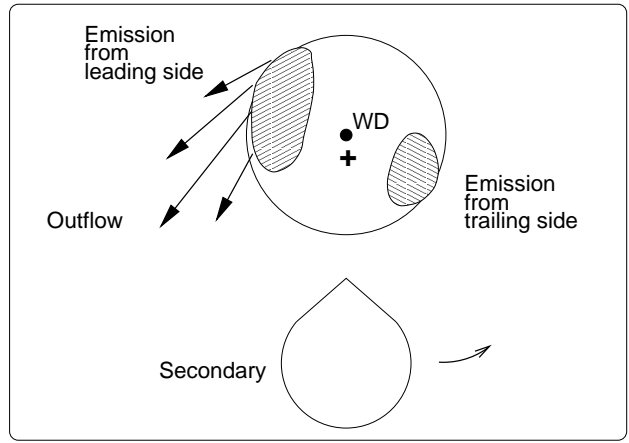}}
\caption{\label{model} The sketch visualises our model of the
emission sources in RR\,Pic and the outflow coming from the 
emission region on the leading side of the disc. During a complete
orbit of the binary rotation, the projected velocity of the outflow 
changes between negative (flow towards the observer) and positive
(flow away from observer) velocities.
 }
\end{figure}

A different explanation is that the high velocity gas  
does not stay inside the system but
is actually ejected in the form of an asymmetric outflow, 
with maximum projected velocity of 1200\,km\,s$^{-1}$.
Taking into account the proposed inclination $i=65^\circ$ 
\citep{haef+82}
the real outflow velocity has to be around 1350\,km\,s$^{-1}$. 
This would explain the variation of the high velocity component
as a lighthouse effect: the outflow emerges
from one part of the disc only and is swept around with the rotation of the 
system. In this picture, both the high velocity itself and its varying
amplitude are explained in a consistent way. Including the information from
phase $\phi_r = 0.25$ of maximal negative velocity (outflow towards us),
the origin of the outflow coincides with the position of the isolated emission
source on the leading side of the accretion disc. 
See Fig.  \ref{model} for an explanatory sketch.

Additional evidence for
the presence of an outflow in RR\,Pic comes from the mapping 
of its remnant shell,
which shows a bipolar structure in the form of an equatorial ring with
some tails orthogonal to it \citep{gill+98}.
Although this map shows the shell far outside the binary,
the ring might well be fed by the 
asymmetric outflow from the binary. 
The fact that no P\,Cygni profiles 
have been found in IUE spectra of RR\,Pic \citep{rosi+82}
has been interpreted as RR\,Pic having no outflow and
that the shell is already decoupled from the binary. 
The asymmetric outflow can be considered as optically thin 
as it is visible in emission. Any P\,Cygni absorption has hence to be weak,
and can furthermore only be produced during the phase when the 
outflow is directed towards the observer.
It would hence smear out over normal exposure times,
and is therefore not in contradiction with the observations.

\section{Conclusion}
\label{conclusion}

We have shown via a Doppler map 
that two isolated emission sources are present in the disc of RR\,Pic.
Combining the radial velocities with published lightcurves
we come to a rather uncommon interpretation of the photometric eclipse as
an eclipse of the leading side emission source (see Fig. \ref{pic3})
rather than an eclipse of the hot spot as suggested previously 
by \citet{kubi84}.

Putting all the information together, we confirm the working model
of \citet{haef+82} for RR\,Pic in preference to the one of \citet{vogt75}. 
Although, as discussed in section \ref{dd}, the superior conjunction of 
the white dwarf takes place 0.1 earlier in phase, this does not change their
basic interpretation. Warner's (1986) doubts of the model were based on the 
fact that no isolated emission from the leading side 
had been found in other CVs. Meanwhile many systems with this structure
have been found (see \citet{tapp+01} for an overview), and this argument is no 
longer valid.

Although common, no convincing explanation for 
isolated emission sources on the leading side of the disc 
has so far been presented. Among
other possibilities, the presence of spiral shock waves has been discussed
as possible explanation for similar features in other CVs
(see e.g. \citet{stegh01}).
Due to their still enhanced mass transfer, long--period old novae should 
possess extended hot discs, and are therefore indeed candidates for 
possessing spiral shocks \citep{boff01}.
However, although the emission sources in the disc of
RR Pic have a clearly elongated structure, neither time nor 
spectral resolution of our data is high enough to 
confirm this picture for RR\,Pic.

The analysis of the high velocity wings of the H$\alpha$ line
yields two possible explanations for these wings and their variability.
They may originate from a partial disc--overflow, or from an asymmetric outflow
from the accretion disc, i.e. from the emission source on the leading side
of the disc. In this case,
a connection between the outflow of the binary and
the ring--like structure of RR\,Pic's shell seems plausible. 
Deep, high resolution
H$\alpha$ images as well as high resolution spectroscopy of some metal lines
 might help to clarify this picture.

\label{lastpage}
\end{document}